\documentclass{article}
\usepackage[utf8]{inputenc}
\usepackage{authblk}
\usepackage[dvips]{graphicx}
\usepackage{amsmath}
\usepackage{amssymb} 
\usepackage{empheq}

\graphicspath{{noiseimages/}}
\usepackage{xcolor}
\usepackage[T2A]{fontenc}
\usepackage{tikz}
\usepackage{MnSymbol}
\usepackage{hyperref}
\usepackage{cite}
\usepackage{caption}
\usepackage{pgfplots}
\usepackage{mathrsfs} 
\pgfplotsset{compat=1.7}
\usetikzlibrary{intersections, pgfplots.fillbetween}
\usetikzlibrary{decorations.pathmorphing}
\usetikzlibrary{arrows.meta}
\usepackage[mode=buildnew]{standalone}
\usepackage[toc,page]{appendix}
\usepackage{amsmath}
\usepackage{float}
 \usepackage{bm}
\usepackage{a4wide}
\usepackage{bm}

\usepackage[english]{babel}
\usepackage{hyperref}
\definecolor{urlcolor}{HTML}{990000}
\definecolor{linkcolor}{HTML}{005F5F} 
\hypersetup{pdfstartview=FitH, linkcolor=linkcolor,urlcolor=urlcolor, colorlinks=true,citecolor=blue}
\usepackage[page]{appendix}

\author[1,3]{\textbf{D. Diakonov}\footnote{\tt dmitrii.dyakonov@phystech.edu}}

\author[1,2,3,4]{\textbf{A. Morozov} \footnote{\tt morozov@itep.ru
}}

\affil[1]{ \itshape MIPT, Dolgoprudny, 141701, Russia}
 
\affil[2]{\itshape NRC “Kurchatov Institute”, 123182, Moscow, Russia}

\affil[3]{\itshape Institute for Information Transmission Problems, Moscow 127994, Russia}
 
 \affil[4]{\itshape ITEP, Moscow, Russia }

\title{\textcolor{black}{Banana diagrams as functions of geodesic distance}}

\date{}
\begin{document}

 \hspace{\fill} MIPT/TH-21/24 \\

 \hspace{\fill} IITP/TH-22/24\\
 
 \hspace{\fill} ITEP/TH-27/24

\begingroup
{\let\newpage\relax

\maketitle}
\endgroup

\begin{abstract}

 We extend the study of banana diagrams in coordinate representation to the case of curved space-times. If the space is harmonic, the Green functions continue to depend on a single variable -- the geodesic distance. But now this dependence can be somewhat non-trivial. We demonstrate that, like in the flat case, the coordinate differential equations for powers of Green functions can still be expressed as determinants of certain operators. Therefore, not-surprisingly, the coordinate equations remain straightforward -- while their reformulation in terms of momentum integrals and Picard-Fuchs equations can seem problematic. However we show that the Feynman parameter representation can also be generalized, at least for banana diagrams in simple harmonic spaces, so that the Picard-Fuchs equations retain their Euclidean form with just a minor modification. A separate story is the transfer to the case when the Green function essentially depends on several rather than a single argument. In this case, we provide just one example, that the equations are still there, but conceptual issues in the more general case will be discussed elsewhere.

\end{abstract}

\newpage
\tableofcontents

\newpage

\section{Introduction}

Feynman integrals are the fundamental building blocks in various fields of physics. Accurately calculating their values and exploring their analytical properties is crucial for our understanding of physical phenomena. Especially important is revealing of algebraic structures, associated with renormalization group, Bogolubov R-operation and hidden integrable structures \cite{Connes_2000,GERASIMOV_2001, dolotin2008introductionnonlinearalgebra, Morozov_2008}. Majority of considerations so far was devoted to Picard-Fuchs equations, providing links to algebraic geometry and different branches of Galois theory \cite{marcolli2009feynmanintegralsmotives, Muller-Stach:2012tgj, Vanhove:2018mto, Lairez:2022zkj, delaCruz:2024xit,Doran:2023yzu,Mishnyakov:2023wpd}. They revealed a rich pattern of differential equations, what implied valuable insights into the analytical characteristics of Feynman diagrams. Much less attention was given to examination of Feynman integrals in configuration space — despite in many respects this is more natural and straightforward. Only recently the simplest banana or sunrise diagrams were analyzed in this way
\cite{Mishnyakov:2024rmb, Cacciatori:2023tzp}, see also \cite{Mishnyakov:2024xjz} for futher directions. The main advantage of banana diagrams is that in configuration space they are just powers of the Green function, without any integrals — and the problem is just to obtain an equation for the power of a function from the one for the function itself. It is not quite simple, still possible.

In this paper we attract attention to another side of the story. Most consideration so far were in Minkowski space-time, where differential equations can be constructed in both coordinate and momentum space. It is in the latter case that integrals and Picard-Fuchs equations for their dependence on parameters emerge — and their connection to equations in configuration space is not fully obvious even in banana case \cite{Mishnyakov:2024rmb }. However, what happens if we switch from the flat space-time to a curved one? In curved space the linear momentum representation is not available because of the lack of translation invariance, and it should be substituted by complete harmonic analysis, which is far more complicated. Thus, in the curved space-time, it is much more practical to study differential equations for Feynman integrals in position space, which are much closer to those in the flat case. Only after that it is reasonable to attempt a return to momentum space.

In this work, we will demonstrate that if the Green function $G(\sigma)$ is a function of one variable $\sigma$(the geodesic distance) and solves the homogeneous Klein-Gordon equation, then the position space differential equations for banana diagrams (for the $n$-th power of $G$) can be represented in a simple form in any dimensional harmonic curved space as the determinant of a certain operator:
\begin{gather}
\label{detDG^n=0}
 \det( \hat{D}_n) G^n(\sigma) = 0,
\end{gather}
for $\sigma > 0$. If the Green function solves the inhomogeneous Klein-Gordon equation, then the right-hand side of equation \eqref{detDG^n=0} is $\delta$-like singular. For a relevant discussion see \cite{Mishnyakov:2023wpd}. For alternative discussions of one and two-loop banana diagrams in curved space (dS and AdS), see \cite{Cacciatori:2024zbe,Cacciatori:2024zrv}.

Let us note that in a time-dependent background, the banana diagram becomes more complicated since one should apply the Keldysh-Schwinger diagram technique rather than Feynman. For example, the de Sitter space is harmonic, and the Green function is a function of geodesic distance, but loop corrections in different patches might break or respect de Sitter isometries \cite{Akhmedov:2022uug}. Therefore, our discussion touches upon the situations where one can apply the Feynman diagram technique.

For general curved space, Picard-Fuchs equations become problematic. However, for simple harmonic space, the Green function looks the same as in Euclidean space. Due to this, we generalize the Feynman parameter representation of the banana diagram for simple harmonic space, and as a result, we conclude that the Picard-Fuchs equation is the same as in Euclidean space.

At the end of this work, we will discuss the potential for generalizing the entire narrative to less trivial scenarios where the Green's function depends on two variables.

This paper is organized as follows: In Section 2, we provide a brief overview of harmonic spaces and explicitly demonstrate that the maximally symmetric space and the complex Grassmannian are harmonic spaces. In Section 3, we rewrite the Klein-Gordon equation in terms of geodesic distance for harmonic space. In Section 4, we present the $\Lambda$-formalism, in which the Klein-Gordon equation has the same form for general harmonic space. In Section 5, we demonstrate that the position space equations for Banana Feynman diagrams can be represented as the determinant of some operator; this structure does not depend on the dimension or type of harmonic space. In Section 6, we generalize the Feynman parameter representation of the banana diagram for simple harmonic space. In Section 7, we discuss the position space differential equation in the scenario where the Green function depends on two variables.
\section{Harmonic space}
In this section, we discuss some examples of harmonic spaces. These spaces allow us to rewrite the Klein-Gordon equation in terms of geodesic distance, which we will further elaborate on in the next section. The definition of harmonic spaces is that the laplacian of the geodesic distance in these spaces is a function of the geodesic distance:

\begin{align}
\label{laplace of geod dist}
 \triangle \sigma =p(\sigma).
\end{align}

One of the key properties of harmonic spaces is that they are Einsteinian \cite{petrov1969einstein,schouten2013ricci}:
\begin{align}
\label{metric is einstainian}
 R_{\mu\nu} = \kappa g_{\mu\nu}.
\end{align}

It was proven in \cite{10.4310/jdg/1214444087} that any space covered by $\mathbb{R}^n$ or isometric to a rank one symmetric space is harmonic. In the compact case, these spaces include the sphere $\mathbb{S}^n$, the real projective space $\mathbb{R}P^n$, the complex projective space $\mathbb{C}P^n$, the quaternionic projective space $\mathbb{H}P^n$, and the Cayley projective plane $\mathbb{O}P^2$; for more details see \cite{Gilkey}. The negative curvature or pseudo-Riemannian analogs of the listed spaces are also harmonic, as they can be obtained through analytical continuation. The classification of harmonic spaces is not yet complete, and not all such spaces are known.

We will further explore in detail some examples of harmonic spaces, including simple harmonic spaces, maximally symmetric spaces, complex projective spaces, and complex Grassmannians.

\subsection{Simple harmonic space} 
Among harmonic spaces, there is a specific category known as simple harmonic spaces (\(SH\)), for which the Laplacian of the geodesic distance appears the same as in $d$-dimentional flat space:
\begin{align}
\label{simply harm def}
 \triangle \sigma = (d-1) \sigma^{-1}.
\end{align}
As an example of a simple harmonic space, one can consider the following space \cite{Walker}:
\begin{align}
 ds^2 = (x^2 dx^1 - x^1 dx^2)^2 + 2 d x^1 d x^3 + 2 d x^2 d x^4.
\end{align}
This space is Ricci flat (\(R_{\mu\nu}=0\)), but it has a non-zero component of the curvature tensor (\(R_{1212}=-3\)). The geodesic distance has a form similar to the geodesic distance in Euclidean space:
\begin{gather}
 \sigma(x,y) = \sqrt{(y_1 x_2 - y_2 x_1)^2 + 2(x_1 - y_1)(x_3 - y_3) + 2(x_2 - y_2)(x_4 - y_4)} = \\= \nonumber \sqrt{g_{\mu\nu}(x^\mu - y^\mu)(x^\nu - y^\nu)},
\end{gather}
satisfying relation \eqref{simply harm def}.

It should also be noted that the product of simple harmonic spaces \(M^{\sum_i d_i} = \prod_{i=1}^n M^{d_i}_i\) is a simple harmonic space, where $d_i$ is the dimension of the space: $d_i=dim\left(M^{d_i}_i\right)$. The geodesic distance in this space, \(\sigma = \sqrt{\sum_i \sigma_i^2}\), where \(\sigma_i\) is the geodesic distance on \(M_i^{d_i}\), satisfies the definition \eqref{simply harm def}:
\begin{gather}
 \triangle \sigma = \sum_i \triangle_i \sqrt{\sum_k \sigma_k^2} = \sum_i \left( \frac{1 + \sigma_i \triangle_i \sigma_i }{\sigma} - \frac{\sigma_i \sigma_i }{\sigma^3} \right) = \\= \nonumber \left(n + \sum_{i=1}^n (d_i - 1) - 1\right) \frac{1}{\sigma} = \nonumber \left(\sum_{i=1}^n d_i - 1\right)\frac{1}{\sigma},
\end{gather}
where we use the fact that the vector \(\nabla_\mu \sigma\) has unit length \eqref{unit vec}. At the same time, it is clear that this is not true for non-simple harmonic spaces since \(\triangle \sigma \nsim \frac{1}{\sigma}\). As an example, we consider at the end of the next subsection the product of two spheres \(S^n \times S^m\).

\subsection{Maximally Symmetric Space}

Maximally symmetric spaces have the following embedding:
\begin{align}
 z^2 + \eta_{\mu\nu} x^\mu x^\nu = r^2
\end{align}
into an \(d+1\) dimensional space with general signature \(\eta_{\mu\nu} = \text{diag}(\pm 1, \pm 1, \ldots, \pm 1)\). The metric is given by:
\begin{align}
\label{metS}
 g_{\mu\nu} = \eta_{\mu\nu} + \frac{x^\alpha x^\beta \eta_{\alpha\mu} \eta_{\beta\nu}}{r^2 - \eta_{\mu\nu} x^\mu x^\nu}.
\end{align}
Using the Christoffel symbols \(\Gamma^\mu_{\nu\rho} = \frac{1}{r^2} x^\mu g_{\nu\rho}\) and the normalization condition \(g_{\nu\rho} \frac{\partial x^\nu}{\partial \tau} \frac{\partial x^\rho}{\partial \tau} = 1\), one can obtain the geodesic equation:
\begin{align}
 \frac{\partial^2 x^\mu}{\partial \tau^2} + \frac{1}{r^2} x^\mu = 0.
\end{align}
The solution of the geodesic equation has the following form:
\begin{align}
 x^\mu = r e^\mu \sin\left(\frac{\tau}{r}\right),
\end{align}
with initial conditions \(x^\mu(\tau = 0) = 0\) and \(\frac{\partial x^\mu}{\partial \tau}(\tau = 0) = e^\mu\), where \(e^\mu\) is a tangent vector with unit length \(g_{\mu\nu}(\tau = 0) e^\mu e^\nu = 1\). Hence, we can conclude that the geodesic distance is defined by the angle between two points on a sphere:
\begin{align}
 \sigma = r \arccos\left(\frac{z}{r}\right) = r \arccos\left(\frac{\sqrt{r^2 - \eta_{\mu\nu} x^\mu x^\nu}}{r}\right).
\end{align}

The Laplacian operator is given by:
\begin{gather}
 \triangle = \nabla_\mu \nabla^\mu = g^{\mu\nu} \partial_\mu \partial_\nu - \Gamma^\mu_{\nu\rho} g^{\nu\rho} \partial_\mu 
 = g^{\mu\nu} \partial_\mu \partial_\nu - \frac{d}{r^2} x^\mu \partial_\mu,
\end{gather}
where the inverse metric is defined as follows:
\begin{align}
 g^{\mu\nu} = \eta^{\mu\nu} - \frac{x^\mu x^\nu}{r^2}.
\end{align}
Since the metric tensor depends only on the tensor \(\eta^{\mu\nu}\) and \(x^\mu x^\nu\), and the second term in the Laplacian is the dilation operator \(x^\mu \partial_\mu\), and given that the geodesic distance is a function of \(\eta_{\mu\nu} x^\mu x^\nu\), it is clear that the Laplacian of the geodesic distance is a function of the geodesic distance:
\begin{align}
\label{lap MAX sym}
 \triangle \sigma = (d-1) \frac{1}{r} \cot \left(\frac{\sigma}{r}\right).
\end{align}

Now, let us show that the product of two maximally symmetric spaces is not a harmonic space. For example, for the product of two spheres \(S^n \times S^m\), one can show that the geodesic distance is given by \(\sigma = \sqrt{\sigma_1^2 + \sigma_2^2}\), where \(\sigma_1\) and \(\sigma_2\) are the geodesic distances on the first and second spheres, respectively. Hence, the Laplacian of the geodesic distance is not a function of the geodesic distance:
\begin{gather}
 \triangle \sigma = \triangle_1 \sqrt{\sigma_1^2 + \sigma_2^2} + \triangle_2 \sqrt{\sigma_1^2 + \sigma_2^2} = \frac{2}{\sigma} + \frac{\sigma_1 \triangle_1 \sigma_1 + \sigma_2 \triangle_2 \sigma_2}{\sigma} = \\= \nonumber 
 \frac{2}{\sigma} + \frac{(n - 1) \sigma_1 \frac{1}{r_1} \cot \left(\frac{\sigma_1}{r_1}\right) + (m - 1) \sigma_2 \frac{1}{r_2} \cot \left(\frac{\sigma_2}{r_2}\right)}{\sigma},
\end{gather}
where we use \eqref{lap MAX sym} and \eqref{unit vec}.

\subsection{$ \mathbb{CP}^n$}

The complex projective space $( \mathbb{CP}^n)$ points label the lines through the origin of a complex $(n+1)$ dimensional complex space. This space is a homogeneous space and is a special case of a complex Grassmannian $(\mathbb{GR}^n(1,n+1))$, whose general case we will consider in the next subsection. The geodesic distance is given by the Hermitian angle between two complex lines:
\begin{align}
 \sigma = \arccos\left(\sqrt{\frac{(Z \cdot \bar{W}) (\bar{Z}\cdot W)}{ (Z \cdot \bar{Z}) (W \cdot \bar{W})}}\right) = \arccos\left(\sqrt{\frac{(1+z \cdot \bar{w}) (1+\bar{z}\cdot w)}{ (1+z \cdot \bar{z}) (1+\bar{w}\cdot w)}}\right),
\end{align}
where \(Z=(Z^0,Z^1,...,Z^n)\) are homogeneous coordinates on \(\mathbb{C}^n\) and \(z_i=\frac{Z^i}{Z^0}\) are the affine coordinates. This distance function is invariant under rescaling \(Z \to \lambda_Z Z\) and \(W \to \lambda_W W\), where \(\lambda_{Z,W} \in \mathbb{C}\).

Using Syngle's formula \eqref{Synge}, one can obtain the metric of space from the definition of geodesic distance:
\begin{align}
 g_{\mu \bar{\nu}} = -\frac{1}{2} \partial_\mu^{(z)} \partial_{\bar{\nu}}^{(\bar{w})} \sigma^2 \Bigg|_{w=z} = \frac{(1+|z|^2)\delta_{\mu \bar{\nu}}-\bar{z}_\mu z_{\nu}}{ (1+|z|^2)^2}.
\end{align}
This is the Fubini-Study metric and it has a Kähler form:
\begin{align}
\label{kahler}
 g_{\mu \bar{\nu}} = \partial_{\mu} \bar{\partial}_{\bar{\nu}} \log\left( 1+z \cdot \bar{z} \right),
\end{align}
due to which the Laplacian does not depend on the Christoffel symbol and is given by:
\begin{align}
\label{lapCP}
 \triangle = 2 g^{\mu \bar{\nu}} \partial_{\mu} \bar{\partial}_{\bar{\nu}}.
\end{align}
To show that the Laplacian of the geodesic distance is a function of the geodesic distance, let us consider the following relation:
\begin{align}
 2 \log(\cos(\sigma)) = \log (1+z \cdot \bar{w}) + \log(1+\bar{z}\cdot w) - \log(1+z \cdot \bar{z}) - \log (1+\bar{w}\cdot w)
\end{align}
Then, using the definition of Laplacian \eqref{lapCP} and the Kähler form of the metric \eqref{kahler}, one can obtain:
\begin{align}
 \triangle 2 \log(\cos(\sigma)) = -2 g^{\mu \bar{\nu}} \partial_{\mu} \bar{\partial}_{\bar{\nu}} \log(1+z \cdot \bar{z}) = -2 \dim [\mathbb{CP}^n].
\end{align}
Hence, using that the vector \(\nabla_\mu \sigma\) has unit length \eqref{unit vec}, we can obtain that the complex projective space is harmonic:
\begin{align}
\triangle \sigma = \left(\dim [\mathbb{CP}^n]-1\right) \cot(\sigma) - \tan(\sigma).
\end{align}

\subsection{Grassmannian}
The Grassmannian \(\mathbb{GR}(k,n)\) is a natural generalization of \( \mathbb{CP}^n\), with the points labeling the \(k\)-planes through the origin of an \(n\)-dimensional complex space. The Grassmannian can be parameterized by Pontrjagin coordinates \(\left(\mathcal{Z}_{ia},\bar{\mathcal{Z}}_{ia}\right)\), where the geodesic distance is given by \cite{gras}:
\begin{align}
 \cos(\sigma) = \sqrt{\frac{\det\left(1_{ab} + \mathcal{Z}_{ia} \bar{\mathcal{W}}_{ib}\right) \det\left(1_{ab} + \bar{ \mathcal{Z}}_{ia} \mathcal{W}_{ib}\right)}{\det\left(1_{ab} + \mathcal{Z}_{ia} \bar{ \mathcal{Z}}_{ib}\right) \det\left(1_{ab} + \mathcal{W}_{ia} \bar{ \mathcal{W}}_{ib}\right)}}.
\end{align}

Using Syngle's formula \eqref{Synge} again, one can obtain the metric:
\begin{align}
 ds^2 = \operatorname{Tr}\left[(1 + \mathcal{Z} \bar{\mathcal{Z}} )^{-1} d \mathcal{Z} d \bar{\mathcal{Z}} - (1 + \mathcal{Z} \bar{ \mathcal{Z}} )^{-1} \mathcal{Z} d\bar{ \mathcal{Z}} (1 + \mathcal{Z} \bar{\mathcal{Z}} )^{-1} d \mathcal{Z} \bar{\mathcal{Z}}\right],
\end{align}
which can also be obtained directly from the Kähler potential  \cite{Morozov:1984ad}:
\begin{align}
 F = \log \det\left(1_{ab} + \mathcal{Z}_{ia} \bar{ \mathcal{Z}}_{ib}\right),
\end{align}
therefore \(\sigma\) is indeed the geodesic distance. Let us note that since $\mathcal{Z}$ is a rectangular matrix, all expressions are invariant under transposition, i.e., $ \mathcal{Z}\to \mathcal{Z}^T$. For example, the Kähler potential can be written in the form:
\begin{align}
 F = \log \det\left(1_{i j} + \mathcal{Z}_{ia} \bar{\mathcal{Z}}_{j a}\right).
\end{align}

Since the metric of the Grassmannian is Kähler, the Laplacian has a simple form:
\begin{align}
 \triangle = 2 g^{(ia) \ \bar{(jb)}} \partial_{ia} \bar{\partial}_{ \bar{jb}},
\end{align}
similar to the previous section, the Laplacian of the geodesic distance can be obtained from the relation:
\begin{align}
 \triangle 2 \log(\cos(\sigma)) = -2 g^{(ia) \ \bar{(jb)}} \partial_{ia} \bar{\partial}_{ \bar{jb}} \log \det\left( 1_{ab} + \mathcal{Z}_{ia} \bar{ \mathcal{Z}}_{ib} \right) = -2 \dim\left[ \mathbb{GR}(k,n)\right].
\end{align}
Hence, using that the vector \(\nabla_\mu \sigma\) has unit length \eqref{unit vec}, we get that the complex Grassmannian manifold is harmonic:
\begin{align}
\triangle \sigma = \left(\dim\left[\mathbb{GR}(k,n) \right] - 1\right) \cot(\sigma) - \tan(\sigma).
\end{align}
Note also that this example does not appear in any literature known to us.

\section{Klein-Gordon equation in terms of geodesic distance}

 In this section, we will show that the Klein-Gordon equation for a massive non-minimally coupled scalar field theory can be rewritten in terms of geodesic distance. We start with Coulomb's law. In $(d > 2)$ dimensional flat space, Coulomb's potential is given by:
\begin{align}
\label{flat Coulomb's potential}
 \varphi \sim \sigma^{2-d},
\end{align}
where \(\sigma= \sqrt{x_i x^i}\) is a geodesic distance. The potential \(\varphi\) solves the differential equation:
\begin{align}
\triangle \varphi(\sigma) = 0, \quad \text{for } \quad \sigma > 0,
\end{align}
which can be rewritten only in terms of geodesic distance: 
\begin{align}
 \left( \partial_\sigma^2 + \frac{d-1}{\sigma} \partial_\sigma \right) \varphi(\sigma) = 0.
\end{align}

In the general curved space, Coulomb's potential is a function of two points. If we suppose that the potential is a function of geodesic distance, then we have:
\begin{align}
\triangle \varphi(\sigma) = \varphi''(\sigma) \, \nabla_\mu \sigma \nabla^\mu \sigma + \varphi'(\sigma) \, \triangle \sigma = 0.
\end{align}
Using the fact that the vector \(\nabla_\mu \sigma\) has unit length: \(\nabla_\mu \sigma \nabla^\mu \sigma = 1\), we can derive the condition that the Coulomb potential is a function of geodesic distance if the Laplacian of the geodesic distance is a function of the geodesic distance:
\begin{align}
\label{laplace of geod dist}
 \triangle \sigma = p(\sigma).
\end{align}
Hence, another definition of a harmonic space is that there exists a harmonic function that depends on geodesic distance.

As a result, in general harmonic space, Coulomb's potential is given by: 
\begin{align}
\label{ Coulomb's potential}
 \varphi(\sigma) = C \int d \sigma \, e^{- \int d \sigma \, p(\sigma)},
\end{align}
where \(C\) is some constant.

Now let us consider the homogeneous Klein-Gordon equation for the Green function of a massive non-minimally coupled scalar field theory:
\begin{align}
(\triangle - m^2 - \xi R) G(\sigma) = 0, \quad \text{for} \quad \sigma > 0.
\end{align}
If the Green function depend on a single variable -- the geodesic distance, then in harmonic space, we can rewrite the Klein-Gordon equation only in terms of geodesic distance, since that space is Einsteinian and has a constant Ricci scalar 
\eqref{metric is einstainian} \(\left
(R = \text{const}\right)
\) we can obtain:
\begin{align}
\label{Gin terms of sigma}
G''(\sigma) + G'(\sigma) \triangle \sigma - (m^2 + \xi R) G(\sigma) = 0,
\end{align}
where the Laplacian of geodesic distance is a function of geodesic distance \eqref{laplace of geod dist}.

\section{$\Lambda$ formalism}
In this section, we introduce the $\Lambda$ formalism in which the Klein-Gordon equation for general harmonic space has a simple form. Using this formalism, we will show in the next section that a position space equation for the banana diagram can be written in a compact form.

Let us introduce the following linear differential operator:
\begin{align}
\Lambda = f(\sigma) \partial_\sigma.
\end{align}
The function \(f\) will be defined below. Then using the following relation: 
\begin{gather}
 \Lambda^2 G = f f' G' + f^2 G'',
\end{gather}
one can rewrite the Klein-Gordon equation \eqref{Gin terms of sigma} in terms of $\Lambda$ as follows:
\begin{align}
\Lambda^2 G - \left[ f' - (\triangle \sigma) f \right] \Lambda G - (m^2 + \xi R) f^2 G = 0. 
\end{align}
Since \(f\) is an arbitrary smooth function, we can choose it to cancel the linear term in $\Lambda$ in the last equation:
\begin{align}
f' = (\triangle \sigma) f = p(\sigma) f \implies f = e^{\int d \sigma \, \triangle \sigma }.
\end{align}
Therefore, if the Green function depends on the geodesic distance, then the Klein-Gordon equation in harmonic space can be written in a simple form:
\begin{align}
\label{KGLambda}
\boxed{\left( \Lambda^2 - \lambda^2 \right) G(\sigma) = 0},
\end{align}
where:
\begin{align}
\Lambda = e^{\int d \sigma \, \triangle \sigma } \partial_\sigma
 \quad \text{and} \quad \lambda = \sqrt{m^2 + \xi R} e^{\int d \sigma \, \triangle \sigma }. 
\end{align}
For example, in an $n$-dimensional Minkowski space, we have:
\begin{align}
\Lambda = \sigma^{d-1} \partial_\sigma
\quad \text{and} \quad \lambda = m \sigma^{d-1},
\end{align}
where $f(\sigma) = \sigma^{d-1}$ is a power function. However, for more general symmetric spaces (when they exist), it can be quite complicated; for more details, see Table \ref{demo-table}.

It is worth noting that to derive \eqref{KGLambda}, we use the fact that the Green function depends on the geodesic distance, and not on an arbitrary function of the geodesic distance.

A similar approach in Minkowski space is discussed in \cite{Mishnyakov:2023sly, Mishnyakov:2024rmb, Mishnyakov:2023wpd}, where the authors choose \(\Lambda = \sigma \partial_{\sigma}\) as a dilatation operator. In our case, the operator \(\Lambda\) for an arbitrary harmonic space is chosen so that when acting on Coulomb's potential \eqref{ Coulomb's potential} it gives a constant:
\begin{align}
\Lambda \varphi = C \quad \implies f = \frac{C}{\varphi'}.
\end{align}
We list the explicit examples of the Laplacian of geodesic distance \(\triangle \sigma\), Coulomb's potential \(\varphi(\sigma)\), and the function \(f(\sigma)\) that define the \(\Lambda\)-operator in Table \ref{demo-table} for a few positive curvature harmonic spaces. For negative curvature, one should replace trigonometric functions with hyperbolic functions \cite{Gilkey}.
 \begin{table}[!h]
\begin{center}
\caption{}
\label{demo-table}
\renewcommand*{\arraystretch}{1}
\begin{tabular}[2]{ |c|c|c|c|c| } 
 \hline
$\mathbb{M}$ & dim$(\mathbb{M})$ & $\triangle \sigma $ or $p(\sigma)$& Coulomb's potential:$\varphi\sim$ & $f(\sigma)$\\ 
\hline
 $SH_n$ & $n$ &$(n-1)\sigma^{-1}$& $\sigma^{2-n}$ & $\sigma^{n-1}$ \\
 \hline
$ \mathbb{S}^n, \ \mathbb{RP}^n$& $n$ & $(n-1) \text{ctg}(\sigma)$& $\int d \sigma \sin(\sigma)^{1-n}$& $\sin^{n-1}(\sigma)$ \\
 \hline
 $\mathbb{CP}^n $& $2n$ & $(2n-1) \text{ctg}(\sigma)-\text{tg}(\sigma)$& $\int d \sigma \sin^{1-2 n}(\sigma) \cos^{-1}(\sigma)$ & $\sin ^{2 n-1}(\sigma) \cos(\sigma)$\\
 \hline
 $\mathbb{GR}(k,n)$ & $d=2 k (n-k)$& $(d-1) \text{ctg}(\sigma)-\text{tg}(\sigma)$& $\int d \sigma \sin^{1- d}(\sigma) \cos^{-1}(\sigma)$ & $\sin ^{ d-1} (\sigma) \cos(\sigma)$\\
 
 \hline
 $\mathbb{HP}^n$ & $4n$ & $(4n-1) \text{ctg}(\sigma)-3\text{tg}(\sigma)$ & $\int d \sigma \sin^{1-4 n}(\sigma) \cos^{-3}(\sigma)$& $\sin^{4 n-1}(\sigma) \cos^{3}(\sigma)$\\
 \hline
 $\mathbb{OP}^2$ & $16$ & $15 \text{ctg}(\sigma)-7 \text{tg}(\sigma) $ & $\int d \sigma \sin^{-15}(\sigma)\cos^{-7}(\sigma) $& $\sin^{15}(\sigma)\cos^{7}(\sigma)$\\
 \hline
 \end{tabular}
\end{center}
\end{table}

\section{Differential equation for banana diagram}
Now, we are going to derive the differential equation for the banana diagram \( B_n = G^n \). In this section, we consider the case when the Green's function depends on one variable (geodesic distance). To achieve this, let us introduce a set of differential operators \( \{O_k\} \) that act on the function \( B_n \) as follows:
\begin{align}
 O_k B_n = B_n^{(k)} (\Lambda G)^k,
\end{align}
where \( B_n^{(k)} = \frac{d^k}{d G^k} B_n \). For example, for \( k=0 \) and \( k=1 \):
\begin{gather}
O_0 B_n = 1 \cdot B_n \quad \text{and} \quad O_1 B_n = \Lambda B_n = B_n^{(1)} \Lambda G.
\end{gather}

To construct the operator \( O_k \), let us apply the operator \(\Lambda\) to \( O_{k-1} B_n \). Then, using \eqref{KGLambda} and the relation \( B_n^{(k)} G = (n-k+1) B_n^{(k-1)} \), we obtain the recurrence relation:
\begin{align}
\label{recO}
 O_k = \Lambda O_{k-1} - (k-1)(n-k+2) \lambda^2 O_{k-2}.
\end{align}
This relation allows us to determine an operator for any given number \( k \). The chain ends at step \( n+1 \) where \( O_{n+1} B_n = 0 \) because \( B_n^{(n+1)} = 0 \).

The recurrence relation \eqref{recO} is similar to the recurrence relation for determinant of a tridiagonal matrix:
\begin{align}
 M_k = a_k M_{k-1} - c_{k-1} b_{k-1} M_{k-1}
\end{align}
where:
\begin{align}
 M_k = \det \begin{pmatrix}
a_1 & b_1 & 0 & ... & 0 & 0 \\
c_1 & a_2 & b_2 & ... & 0 & 0 \\
0 & c_2 & a_3 & ... & 0 & 0 \\
 & & \cdot & \cdot &\cdot & \\
0 & 0 & 0 & ... & a_{k-1} & b_{k-1} \\
0 & 0 & 0 & ... & c_{k-1} & a_k \\
\end{pmatrix}.
\end{align}

Hence, each operator \( O_k \) can be represented as the determinant of a tridiagonal matrix. Therefore, the differential equation for the banana diagram \( O_{n+1} B_n = 0 \) can be expressed as follows:
\begin{align}
\label{DetLl}
\boxed{
 \det \begin{pmatrix}
\Lambda & c_1 \lambda & 0 & ... & 0 & 0 \\
c_1 \lambda & \Lambda & c_2 \lambda & ... & 0 & 0 \\
0 & c_2 \lambda & \Lambda & ... & 0 & 0 \\
 & & \cdot & \cdot &\cdot & \\
0 & 0 & 0 & ... & \Lambda & c_{n} \lambda \\
0 & 0 & 0 & ... & c_{n} \lambda & \Lambda \\
\end{pmatrix} G^n=0},
\end{align}
where \( c_k = \sqrt{k(n-k+1)} \). In this formula, the operators are ordered so that the operators from the line above act first. It is worth noting that the structure of this differential equation does not depend on the dimension or the type of harmonic space.

For \( n=1 \), the determinant gives the Klein-Gordon equation \eqref{KGLambda}: 
\begin{align}
 \hat{D}_1 = \det \begin{pmatrix}
\Lambda & \lambda \\
\lambda & \Lambda
\end{pmatrix} = \Lambda^2 - \lambda^2.
\end{align}
For \( n=2 \), one can obtain: 
\begin{align}
 \hat{D}_2 = \det \begin{pmatrix}
\Lambda & \sqrt{2} \lambda & 0 \\
\sqrt{2} \lambda & \Lambda & \sqrt{2} \lambda \\
0 & \sqrt{2} \lambda & \Lambda 
\end{pmatrix} = \Lambda^3 - 2\Lambda \lambda^2 - 2\lambda^2 \Lambda,
\end{align}
and so on.
\section{PF for simle harmonic space}
In Euclidean space, one can use momentum representation to derive the Feynman parameter representation of the Banana diagram \cite{Weinzierl:2022eaz}:
\begin{align}
 \label{momentum repres}
 \widetilde{B}_{n}(p^2) = \int_0^\infty \left[\prod_{i=1}^n d \alpha_i\right] \delta \left(1-\sum_{i=1}^n \alpha_i\right) \frac{U^{\frac{n(2-d)}{2}} (\alpha)}{\left(\alpha_1 \cdots \alpha_n p^2 - m^2 U(\alpha) \sum_{i=1}^n \alpha_i\right)^{1+\frac{(n-1)(2-d)}{2}}},
\end{align}
where $\alpha$ is a Feynman parameter and \( U(\alpha) = \prod_{i=1}^n \alpha_i \sum_{j=1}^n \alpha_j^{-1} \). In this case, banana integrals satisfy Picard-Fuchs equations \cite{Muller-Stach:2012tgj}:
\begin{align}
\label{PF}
 \hat{PF}_{p^2} \cdot \widetilde{B}_{n}(p^2) = 0.
\end{align}

The derivation of these equations is achieved directly from the momentum representation \eqref{momentum repres} and provides a simpler form and much lower order compared to deriving it through a Fourier transform from the configuration space equation \eqref{DetLl}, for more details, see \cite{Mishnyakov:2024rmb}. In curved space, where differential equations for Feynman integrals exist in configuration space \eqref{DetLl}, linear momentum space is absent. However, there is complete harmonic analysis; using that, we will show that the Feynman parameter representation of the Banana diagram can be obtained in the harmonic space.

The Klein-Gordon equation for harmonic spaces takes the form:
\begin{gather}
 (\triangle - m^2 - \xi R) G(\sigma) = G''(\sigma) \nabla_\mu \sigma \nabla^\mu \sigma + G'(\sigma) \triangle \sigma - (m^2 + \xi R) G(\sigma) = \\ \nonumber
 = G''(\sigma) + (d-1) \sigma^{-1} G'(\sigma) - (m^2 + \xi R) G(\sigma) = 0, \quad \text{for} \quad \sigma > 0.
\end{gather}

This equation is similar to the differential equation for the Green's function in Euclidean space with a shifted mass: $m^2 \to m^2 + \xi R$ where the Ricci scalar is constant. Therefore, the Green's function in simple harmonic space is given by:
\begin{align}
 G(\sigma) \sim \int_0^\infty \frac{ds}{s^{\frac{d}{2}}} e^{-\frac{\sigma^2}{4s} - (m^2 + \xi R) s}.
\end{align}

The heat kernel solves the following differential equation:
\begin{align}
 (\triangle - m^2 - \xi R) K(s, \sigma) = \partial_s K(s, \sigma).
\end{align}
The solution can be expressed as:
\begin{align}
\label{heat}
 K(s, \sigma) = \sumint_i \ e^{-s\lambda_i} \phi_i(x) \phi_i^*(y) \sim \frac{1}{s^{\frac{d}{2}}} e^{-\frac{\sigma^2}{4s} - (m^2 + \xi R) s},
\end{align}
where $\phi_i(x)$ and $\lambda_i$ are the eigenfunctions and eigenvalues of the Klein-Gordon equation:
\begin{align}
 (\triangle - m^2 - \xi R) \phi_i(x) = -\lambda_i \phi(x),
\end{align}
and the index $i$ can be discrete and continuous. Eigenfunctions are orthonormal and complete:
\begin{align}
 \int d^d x \sqrt{g} \phi_i(x) \phi_i^*(x) = \delta_{ij} \quad \text{and} \quad \sumint_i \phi_i(x) \phi_i^*(y) = \frac{\delta(x-y)}{\sqrt{g}}.
\end{align}

Now, let us consider multi-loop banana diagrams:
\begin{align}
 B_n = G^n(\sigma) \sim \int_0^\infty \left[\prod_{k=1}^n \frac{ds_k}{s_k^{\frac{d}{2}}}\right] e^{-\frac{\sigma^2}{4} \left( \sum_{k=1}^n \frac{1}{s_k} \right) } e^{- (m^2 + \xi R) (\sum_{k=1}^n s_k)}.
\end{align}
Using \eqref{heat}, we can express the first exponential term in the last equation in the form:
\begin{align}
 e^{-\frac{\sigma^2}{4 \left( \sum_{k=1}^n \frac{1}{s_k} \right)^{-1}}} \sim \left( \sum_{k=1}^n \frac{1}{s_k} \right)^{-d/2} \sumint_i \ e^{-\left( \sum_{k=1}^n \frac{1}{s_k} \right)^{-1} (\lambda_i - m^2 - \xi R)} \phi_i(x) \phi_i^*(y).
\end{align}
Therefore, we can rewrite the equation for the banana diagram in terms of mode expansion:
\begin{gather}
 B_n(x, y) \sim \\ \nonumber \sim
 \int_0^\infty \left[\prod_{k=1}^n \frac{ds_k}{s_k^{\frac{d}{2}}}\right]
 \left( \sum_{k=1}^n \frac{1}{s_k} \right)^{-\frac{d}{2}} \sumint_i \ e^{-\left( \sum_{k=1}^n \frac{1}{s_k} \right)^{-1} (\lambda_i - m^2 - \xi R) - (m^2 + \xi R) (\sum_{k=1}^n s_k)} \phi_i(x) \phi_i^*(y).
\end{gather}

From this, it becomes straightforward to obtain the momentum representation analog of the banana diagram:
\begin{gather}
 \delta_{ij} \widetilde{B}_n(\lambda_i) = \int d^d x \int d^d y \ \phi_i(x) \phi_j^*(y) B_n(x, y) 
 \sim \\ \nonumber \sim 
 \delta_{ij} \int_0^\infty \left[\prod_{k=1}^n \frac{ds_k}{s_k^{\frac{d}{2}}}\right]
 \left( \sum_{k=1}^n \frac{1}{s_k} \right)^{-\frac{d}{2}} \ e^{-\left( \sum_{k=1}^n \frac{1}{s_k} \right)^{-1} (\lambda_i - m^2 - \xi R) - (m^2 + \xi R) (\sum_{k=1}^n s_k)}.
\end{gather}

Using the fact that the sum of the Schwinger parameters is non-negative, we can insert a 1 in the form of \( 1 = \int_0^\infty dt \ \delta \left( t - \sum_{i=1}^n s_i \right) \) in last integral. By changing the variables as \( s_i \to t \alpha_i \), one can obtain integrals in terms of the Feynman parameter:

\begin{empheq}[box=\fbox]{gather}
 \widetilde{B}_n(\lambda_i) \sim \\ \nonumber \sim 
 \int_0^\infty \left[\prod_{i=1}^n d \alpha_i\right] \delta \left(1 - \sum_{i=1}^n \alpha_i\right) \frac{U^{\frac{n(2-d)}{2}} (\alpha)}{\left( \alpha_1 \cdots \alpha_n (\lambda_i - m^2 - \xi R) - (m^2 + \xi R) U(\alpha) \sum_{i=1}^n \alpha_i \right)^{1 + \frac{(n-1)(2-d)}{2}}}
\end{empheq}

As a result, the Feynman parameter representation of the banana diagram in a simple harmonic space is the same as in Euclidean space with replaced parameters: $p^2 \to \lambda_i - m^2 - \xi R$ and $m^2 \to m^2 + \xi R$. Consequently, the Picard-Fuchs equations are the same as in Euclidean space \eqref{PF}.
 
\section{Function of two variables}
In this section, we consider the case where the Green function depends on two variables. As we will see, there is no simple generalization of the representation of the differential equation in terms of determinants as in \eqref{DetLl}. However, a recurrence relation similar to \eqref{recO} still exists, which can be resolved to obtain a position space differential equation. 

The simplest example is Euclidean space with Dirichlet boundary conditions: $\phi(x)\big|_{x_1=0}=0$. In this case, the Green function depends on two variables:
\begin{align}
 \sigma=\sum_{i=1}^d (x_i-y_i) (x_i-y_i) \quad \text{and} \quad \bar{\sigma}= (x_1+y_1) (x_1+y_1)+ \sum_{i=2}^d (x_i-y_i) (x_i-y_i).
\end{align}

The Green function, which satisfies the Dirichlet boundary condition is given by the expression:
\begin{align}
 G(\sigma,\bar{\sigma})=G(\sigma)-\bar{G}(\bar{\sigma}),
\end{align}
where $G(\sigma)$ and $\bar{G}(\bar{\sigma})$ are the ordinary invariant Green functions that satisfy the following differential equations:
\begin{align}
 (\Lambda^2 -\lambda^2 )G(\sigma)=0, \quad \text{where} \quad \Lambda = f(\sigma) \partial_\sigma
\end{align}
and 
\begin{align}
 (\bar{\Lambda}^2 - \bar{\lambda}^2) \bar{G}(\bar{\sigma})=0, \quad \text{where} \quad \bar{\Lambda} = f(\bar{\sigma}) \partial_{\bar{\sigma}}.
\end{align}

The Green function $G(\sigma, \bar{\sigma})$ then solves the following second-order differential equation: 
\begin{align}
 \left(\frac{1}{\lambda^2}\Lambda^2+ \frac{1}{\bar{\lambda}^2}\bar{\Lambda}^2-1\right) G(\sigma,\bar{\sigma})=0. 
\end{align}
In this scenario, the differential equation for the Green function can be rewritten in terms of the two distances, since $\sigma$ and $\bar{\sigma}$ are geodesic distances. 

Similar to the previous section, there is a set of differential operators $O_n^{k,s}$ such that: 
\begin{align}
 O_n^{k,s} F = F^{(k)} (\Lambda G)^s (\bar{\Lambda}\bar{G})^{k-s}.
\end{align}
Then one can obtain the recurrence relation:
\begin{align}
 \frac{k+2-s}{\lambda^2} \left( \Lambda O^{k+1,s}_{n} - O^{k+2,s+1}_{n} \right) + \frac{s}{\bar{\lambda}^2} \left(\bar{\Lambda} O^{k+1,s-1}_{n} - O^{k+2,s-1}_{n}\right) = s (k+2-s)(n-k) O_n^{k,s-1}
\end{align}
with the conditions: 
\begin{align}
 O^{0,0}_n=1, \quad O_n^{1,0} =\bar{\Lambda}, \quad O^{1,1}_n = \Lambda \quad \text{and} \quad O_{n}^{n+1,s}=0.
\end{align}

In the previous section, the system of differential equations \eqref{recO} was solved recursively by sequential substitution or using the determinant \eqref{DetLl}. However, this system does not yet have a simple and understandable way to solve it for any order. For example, for $n=2$, the system of recurrence relations consists of 10 equations:

\begin{gather}
 O^{1,1}_2 = \Lambda O^{0,0}_2, \quad O^{1,0}_2 = \bar{\Lambda} O^{0,0}_2, \quad O^{2,1}_2 = \Lambda O^{1,0}_2, \quad O^{2,1}_2 = \bar{\Lambda} O^{1,1}_2, \nonumber \\
 \frac{1}{\lambda} \Lambda O^{2,1}_2 + \frac{1}{2 \bar{\lambda}} \bar{\Lambda} O^{2,0}_2 = O^{1,0}_2, \quad \frac{1}{2\lambda} \Lambda O^{2,2}_2 + \frac{1}{\bar{\lambda}} \bar{\Lambda} O^{2,1}_2 = O^{1,1}_2, \nonumber \\
 \frac{1}{\lambda} \Lambda O^{1,1}_2 + \frac{1}{\bar{\lambda}} \bar{\Lambda} O^{1,0}_2 - \frac{1}{\lambda} O^{2,2}_2 - \frac{1}{\bar{\lambda}} O^{2,0}_2 - 2 O^{0,0}_2 = 0, \nonumber \\
 O^{0,0}_2 = 1, \quad \Lambda O^{2,0}_2 = 0 \quad \text{and} \quad \bar{\Lambda} O^{2,2}_2 = 0.
\end{gather}

In this example, there are six number of operators $O_n^{k,s}$. One can resolve this system and obtain the following differential equation:
\begin{align}
\Bigg((\bar{\Lambda}\bar{\lambda})\left[\Lambda^3+3\frac{\lambda}{\bar{\lambda}} \bar{\Lambda}^2 \Lambda - 4\lambda \Lambda\right] 
 + (\Lambda\lambda)\left[\bar{\Lambda}^3 + 3 \frac{\bar{\lambda}}{\lambda} \Lambda^2 \bar{\Lambda} - 4 \bar{\lambda} \bar{\Lambda}\right] - 2 (\bar{\Lambda}\bar{\lambda}) (\Lambda\lambda)\Bigg)B_2 = 0.
\end{align}

Additionally, note that the order of the differential equation is three, which is the same as in the one-variable case \eqref{DetLl}. 
\section{Conclusion}
In this paper, we demonstrate that the position space differential equations for n-banana Feynman diagrams in harmonic space can be expressed concisely as the determinant of a certain operator. The structure of this equation is independent of the type of space being considered, arising from the fact that the Green function depends only on the geodesic distance. For an arbitrary space, the Green function can depend on both points, and the form of the differential equation for n-banana Feynman diagrams is still unknown. In an effort to extend the position-space differential equations for n-banana Feynman diagrams to the case where the Green function is a function of two variables, we examined half of Minkowski spacetime. However, even from this simple example, we notice that the structure of the differential equation becomes significantly more complicated. Hence, it is necessary to develop new methods to generalize the position space differential equation to arbitrary cases.

Using heat kernel analysis, we also conclude that the Picard-Fuchs equations for the banana diagram in simple harmonic space are the same as in Euclidean space, since in that space the Green function is identical to the Euclidean Green function. The generalization of the Feynman parameter representation of the banana diagram and Picard-Fuchs equations for other harmonic spaces \cite{helgason1979differential,Morozov:1984ew} is a subject for further research.
\section*{Acknowledgments}
We would like to thank M. Reva for illuminating discussions. The work was partially funded within the state assignment of the Institute for Information Transmission Problems of RAS. Our work is partly supported by grant RFBR 21-51-46010 ST\_a, by the grants of the Foundation for the Advancement of Theoretical Physics and Mathematics “BASIS”.
\begin{appendices} 

\setcounter{equation}{0}
\renewcommand\theequation{A.\arabic{equation}}
The geodesic distance is defined as follows:
\begin{align}
 \sigma = \int_0^\sigma ds = \int_0^\sigma \sqrt{g_{\mu \nu} dx^\mu dx^\nu} = \int_0^\sigma ds \sqrt{g_{\mu \nu} \frac{dx^\mu}{ds} \frac{dx^\nu}{ds}}.
\end{align}
The small variation to one of the endpoints of the geodesic is given by: 
\begin{align}
 \delta \sigma = -\int_0^\sigma ds \left(\Gamma_{\rho \lambda}^\nu \frac{dx^\rho}{ds} \frac{dx^\lambda}{ds} + \frac{d^2 x^\nu}{ds^2}\right) g_{\mu \nu}\delta x^\mu + \frac{dx_\mu}{ds} \delta x^\mu.
\end{align}
The first term vanishes for geodesics. Hence, the derivative of the geodesic distance with respect to one of its arguments is a vector with unit length:
\begin{align}
\label{unit vec}
 \partial_\mu \sigma = \frac{dx_\mu}{ds} = u_\mu \quad \Rightarrow \partial_\mu \sigma \partial^\mu \sigma = 1.
\end{align}

Let us consider the derivatives of Synge’s world function \( L = \frac{\sigma^2}{2} \):
\begin{align}
 \partial_\mu L = \sigma u_\mu,
\end{align}
The length of the vector is given by:
\begin{align}
 \partial_\mu L \partial^\mu L = 2L.
\end{align}
Taking twice the covariant derivatives of the left-hand and right-hand sides of the last equation, we get \cite{synge1960relativity,DeWitt:1964mxt,ChristensenPhysRevD.14.2490}:
\begin{align}
\left( \nabla_\rho \nabla_\nu \partial_\mu L \right) \partial^\mu L + \left( \nabla_\nu \partial_\mu L \right) \left( \nabla_\rho \partial^\mu L \right) = \nabla_\rho \partial_\nu L.
\end{align}
The coincidence limit of this equation is defined as:
\begin{align}
 A_{\mu \nu} g^{\mu \lambda} A_{\rho \lambda} = A_{\rho \nu},
\end{align}
where we use that at the coincidence limit, \( \partial_\mu L \big|_{y=x} = 0 \) and denote \( A_{\mu\nu} = \lim_{x \to y} \nabla_\nu \partial_\mu L(x,y) \) for derivatives with respect to \( x \). Assuming that \( A_{\mu \nu} \) is an invertible matrix, we get:
\begin{align}
 A_{\mu\nu} = g_{\mu \nu}.
\end{align}

Now, using the fact that at the coincident point, the derivatives of geodesic distance with respect to points \( x \) and \( y \) are related as follows:
\begin{align}
 \partial^{(x)}_\mu \sigma \big|_{y=x} = - \partial^{(y)}_\mu \sigma \big|_{y=x}.
\end{align}
Using that, we can show that the metric of the space can be obtained by using non-covariant derivatives of Synge's world function with respect to the endpoints:
\begin{align}
\label{Synge}
 g_{\mu\nu} = - \frac{1}{2} \partial^{(x)}_\mu \partial^{(y)}_\nu \sigma^2 \big|_{y=x}.
\end{align}
This formula can look similar to expression of Kahler metric through Kahler potential, but in fact it is far more general and true for arbitrary spaces, including those of odd diimension. It reflects the fact that distance and metric are basically the same: one can define and measure the geodesic distance with the help of metric, and one can do just the opposite — define the metric (modulo gauge transformations) at a given point from the geodesic distance in the vicinity of this point. 

For example, for the sphere, the geodesic distance is defined as the angle between two points: \(\sigma = \arccos\left(X_\alpha Y^\alpha\right)\), where \(X^\alpha\) is the embedding coordinate. Hence, using Synge's formula \eqref{Synge}, we can obtain the metric on the sphere:

\begin{align}
\label{Syngle's}
 g_{\mu \nu} = -\frac{1}{2} \partial_\mu^{(x)} \partial_\nu^{(y)} \sigma^2(x,y) \Bigg|_{y=x} = \eta_{\alpha \beta} \partial_\mu X^\alpha \partial_\nu X^\beta.
\end{align}

For the parameterization \(X^\alpha = (z = \sqrt{r^2 - \eta_{\mu \nu} x^\mu x^\nu}, x^i)\), we can obtain the same metric as in \eqref{metS}.

\end{appendices}
 
\bibliographystyle{unsrturl}
\bibliography{bibliography.bib}
\end{document}